\documentclass[final]{siamltex}
\usepackage{amssymb}
\usepackage{showkeys}
\usepackage{graphicx}
\usepackage{amsmath}
\usepackage{amssymb}
\usepackage{graphicx}
\usepackage{epsfig}
\usepackage{labelfig}
\usepackage{amsmath}
\usepackage{color}
\usepackage{afterpage}
\usepackage{verbatim}
\usepackage{umlaut}
\usepackage{afterpage}
\usepackage[small]{caption}

\input{tcilatex}
\begin{document}

\title{ Thermoacoustic tomography with an arbitrary elliptic operator}
\author{Michael V. Klibanov$^{\ast }$}
\thanks{Department of Mathematics and Statistics, University of North
Carolina at Charlotte, Charlotte, NC 28223, USA, email: mklibanv{@}uncc.edu}
\maketitle

\begin{abstract}
Thermoacoustic tomography is a\ term for the inverse problem of determining
of one of initial conditions of a hyperbolic equation from boundary
measurements. In the past publications both stability estimates and
convergent numerical methods for this problem were obtained only under some
restrictive conditions imposed on the principal part of the elliptic
operator. In this paper logarithmic stability estimates are obatined for an
arbitrary variable principal part of that operator. Convergence of the
Quasi-Reversibility Method to the exact solution is also established for
this case. Both complete and incomplete data collection cases are considered.
\end{abstract}



\section{Introduction}

\label{sec:1}

The goal of this paper is to show that logarithmic stability estimates as
well as convergent numerical methods for the inverse problem of determining
an initial condition in a general hyperbolic PDE of the second order can be
obtained without any restrictions on its coefficients, except of some
natural ones. In all previous publications on this topic the principal part
of the elliptic operator was subjected to some restrictive conditions.
Naturally, our stability estimates imply uniqueness. Both complete and
incomplete data collection cases are considered. We assume here that the
data are given on the infinite time interval $t\in \left( 0,\infty \right) .$
Second and third Remarks 2.1 (section 2) justify this assumption. For
brevity, we leave for possible future publications the finest assumptions,
like, e.g. the minimal smoothness, etc.

In thermoacoustic tomography (TAT) a short radio frequency pulse is sent in
a biological tissue \cite{AK,FR}. Some energy is absorbed. It is well known
that malignant legions absorb more energy than healthy ones. Then the tissue
expands and radiates a pressure wave which is the solution of the following
Cauchy problem
\begin{eqnarray}
u_{tt} &=&c^{2}\left( x\right) \Delta u,x\in \mathbb{R}^{3},t>0,  \label{1.0}
\\
u\left( x,0\right) &=&f\left( x\right) ,u_{t}\left( x,0\right) =0.
\label{1.01}
\end{eqnarray}%
The function $u\left( x,t\right) $ is measured by transducers at certain
locations either at the boundary of the medium of interest or outside of
this medium. The function $f\left( x\right) $ characterizes the absorption
of the medium. Hence, if one would know the function $f\left( x\right) $,
then one would know locations of malignant spots. The inverse problem
consists in determining $f\left( x\right) $ using those measurements.

Both stability estimates and convergent numerical methods for the problem of
determining the initial condition $f$ in (\ref{1.01}) are currently known
only under some restrictive conditions imposed on the coefficient $c\left(
x\right) $ (subsection 1.2). In addition, except of the case $c\left(
x\right) \equiv 1$ in \cite{KY}, those numerical methods are known only for
the case of complete data collection, i.e. when boundary measurements are
given at the entire boundary of the domain of interest.

First, we apply a well known analog of the Laplace transform to obtain a
similar inverse problem for a parabolic PDE. Next, previous results of the
author \cite{Kl1,Kl2} are used. In the complete data case the logarithmic
stability estimate follows from \cite{Kl1}. In the case when the data are
given on a hyperplane, we significantly modify the proof of Theorem 1 of
\cite{Kl2}. More precisely, we prove our logarithmic stability estimate for
an integral inequality rather than for the parabolic PDE. We need this
generalization to establish convergence rate of our numerical method.
Results of both publications \cite{Kl1,Kl2} were obtained via Carleman
estimates. In particular, a quite technical non-standard Carleman estimate
was derived in \cite{Kl2}, see Lemma 2.1 in section 2. We refer to \cite{LYZ}
for another logarithmic stability estimate of the initial condition of a
parabolic equation with the self-adjoint operator $L$ in a finite domain. A
Carleman estimate was also used in this reference. An interesting feature of
\cite{LYZ} is that observations are performed on an internal subdomain for
times $t\in \left( \tau ,T\right) $ where $\tau >0.$ In addition, a
numerical method was developed in \cite{LYZ}.

\subsection{Statements of inverse problems}

\label{sec:1.1}

Let $\Omega \subset \left\{ x_{1}>0\right\} $ be a bounded domain with the
boundary $\partial \Omega \in C^{3}$. Let $T>0$. Denote%
\begin{equation*}
Q_{T}=\Omega \times \left( 0,T\right) ,S_{T}=\partial \Omega \times \left(
0,T\right) ,P=\left\{ x_{1}=0\right\} ,P_{T}=P\times \left( 0,T\right) .
\end{equation*}%
Let $k\geq 0$ be an integer and $\alpha \in \left( 0,1\right) $. Below $%
C^{k+\alpha },C^{2k+\alpha ,k+\alpha /2}$ are H\"{o}lder spaces. Consider
the elliptic operator $L$ of the second order with its principal part $%
L_{0}, $
\begin{eqnarray}
Lu &=&\sum\limits_{i,j=1}^{n}a_{i,j}\left( x\right)
u_{x_{i}x_{j}}+\sum\limits_{j=1}^{n}b_{j}\left( x\right)
u_{x_{j}}+b_{0}\left( x\right) u,x\in \mathbb{R}^{n},  \label{1.2} \\
L_{0}u &=&\sum\limits_{i,j=1}^{n}a_{i,j}\left( x\right) u_{x_{i}x_{j}},
\label{1.2_1} \\
a_{i,j} &\in &C^{k+\alpha }\left( \mathbb{R}^{n}\right) \cap C^{1}\left(
\mathbb{R}^{n}\right) ,b_{j},b_{0}\in C^{k+\alpha }\left( \mathbb{R}%
^{n}\right) ,k\geq 2,\alpha \in \left( 0,1\right) ,  \label{1.3} \\
\mu _{1}\left\vert \eta \right\vert ^{2} &\leq
&\sum\limits_{i,j=1}^{n}a_{i,j}\left( x\right) \eta _{i}\eta _{j}\leq \mu
_{2}\left\vert \eta \right\vert ^{2},\forall x,\eta \in \mathbb{R}^{n};\mu
_{1},\mu _{2}=const.>0.  \label{1.4}
\end{eqnarray}%
Let the function $f\left( x\right) $ be such that%
\begin{equation}
f\in C^{p}\left( \mathbb{R}^{n}\right) ,p\geq 3,f\left( x\right) =0,x\in
\mathbb{R}^{n}\diagdown \Omega .  \label{1.5}
\end{equation}%
Consider the following Cauchy problem%
\begin{eqnarray}
u_{tt} &=&Lu,x\in \mathbb{R}^{n},t\in \left( 0,\infty \right) ,  \label{1.6}
\\
u\left( x,0\right) &=&f\left( x\right) ,u_{t}\left( x,0\right) =0.
\label{1.7}
\end{eqnarray}%
We use everywhere below the following assumption.

\textbf{Assumption}. We assume that integers $k\geq 2,p\geq 3$ in (\ref{1.3}%
), (\ref{1.5}), coefficients of the operator $L$ and the initial condition $%
f $ are such that there exists unique solution $u\in C^{3}\left( \mathbb{R}%
^{n}\times \left[ 0,T\right] \right) ,\forall T>0$ of the problem (\ref{1.6}%
), (\ref{1.7}) satisfying
\begin{equation}
\left\Vert u\right\Vert _{C^{3}\left( \mathbb{R}^{n}\times \left[ 0,T\right]
\right) }\leq Be^{dT},\forall T>0,  \label{1.171}
\end{equation}%
where the constants $B=B\left( L\right) >0,d=d\left( L\right) >0$ depend
only from the coefficients of the operator $L$ and an upper estimate $%
\overline{B}$ of the norm $\left\Vert f\right\Vert _{C^{p}\left( \overline{%
\Omega }\right) }\leq \overline{B}.$

Note that (\ref{1.5}) in combination with the finite speed of propagation of
the solution of problem (\ref{1.6}), (\ref{1.7}) guarantee that the function
$u\left( x,t\right) $ has a finite support $\Psi \left( T\right) \subset
\mathbb{R}^{n},\forall t\in \left( 0,T\right) ,\forall T>0$ \cite{Lad}.
Hence, $C^{3}\left( \mathbb{R}^{n}\times \left[ 0,T\right] \right) $ above
is actually the space $C^{3}\left( \overline{\Psi \left( T\right) }\times %
\left[ 0,T\right] \right) .$ Using the classical tool of energy estimates
\cite{Lad}, one can easily find non-restrictive sufficient conditions
imposed on both coefficients of the operator $L$ and the function $f$
guaranteeing the smoothness $u\in C^{3}\left( \mathbb{R}^{n}\times \left[ 0,T%
\right] \right) ,\forall T>0$ as well as (\ref{1.171}). We are not doing
this here for brevity. We consider the following two Inverse Problems.

\textbf{Inverse Problem 1 (IP1).}\emph{\ Suppose that conditions (\ref{1.2}%
)-(\ref{1.5}) and Assumption hold. Let }

$u\in C^{3}\left( \mathbb{R}^{n}\times \left[ 0,T\right] \right) ,\forall
T>0 $\emph{\ be the solution of the problem (\ref{1.6}), (\ref{1.7}). Assume
that the function }$f\left( x\right) $\emph{\ is unknown. Determine this
function, assuming that the following function }$\varphi _{1}\left(
x,t\right) $\emph{\ is known}%
\begin{equation}
u\mid _{S_{\infty }}=\varphi _{1}\left( x,t\right) .  \label{1.8}
\end{equation}

\textbf{Inverse Problem 2 (IP2).} \emph{Suppose that conditions (\ref{1.2})-(%
\ref{1.5}) are Assumption hold. Let }$u\in C^{3}\left( \mathbb{R}^{n}\times %
\left[ 0,T\right] \right) ,\forall T>0$\emph{\ be the solution of the
problem (\ref{1.6}), (\ref{1.7}). Assume that the function }$f\left(
x\right) $\emph{\ is unknown. Determine this function, assuming that the
following function }$\varphi _{2}\left( x,t\right) $\emph{\ is known}%
\begin{equation}
u\mid _{x\in P_{\infty }}=\varphi _{2}\left( x,t\right) .  \label{1.9}
\end{equation}

IP1 has complete data collection, since the function $\varphi _{1}$ is known
at the entire boundary of the domain of interest $\Omega .$ On the other
hand, IP2 represents a special case of incomplete data collection, since $%
\Omega \subset \left\{ x_{1}>0\right\} .$

\subsection{Brief overview of published results}

\label{sec:1.2}

TAT has attracted a significant interest in the past several years. We now
provide a brief overview of published mathematical results for TAT. We refer
to \cite{Kuch} for a review paper. Stability estimates and convergent
numerical methods for an arbitrary time independent principal part $L_{0}$
in (\ref{1.2_1}) were not obtained in the past. Explicit formulas for the
reconstruction of the function $f\left( x\right) $ for IP1 in the case when
in (\ref{1.0}) $c\equiv 1$ are given in a number of publications, see, e.g.
\cite{FPR,FHR,FR,Kuch,KY}. These formulas lead to some stability estimates
as well as to numerical methods with good performances.

Another approach to IP1, IP2 is via analyzing the case when both Dirichlet
and Neumann data are given at $S_{T}$ for IP1 and at $P_{T}$ for IP2. An
elementary, well known and stable procedure of deriving the Neumann
condition from the given Dirichlet condition for both IP1 and IP2 is
described in subsection 2.1 for the parabolic PDE. A very similar procedure
takes place in the hyperbolic case. Consider now IP1. Since a certain norm
of the Neumann boundary condition at $S_{T}$ can be estimated from the above
by another norm of the data $\varphi _{1}\left( x,t\right) $ for $\left(
x,t\right) \in S_{T},$ then the problem of determining the initial condition
$f\left( x\right) $ can be reformulated in a slightly more general setting
as the Cauchy problem for equation (\ref{1.6}) with the lateral Dirichlet
and Neumann data at $S_{T}.$ This problem consists in determining the
function $u\left( x,t\right) $ inside of the time cylinder $Q_{T}$.

We now comment on the Lipschitz stability estimate for that Cauchy problem
with lateral data for the particular case when initial conditions are as in (%
\ref{1.7}). Consider the even extension of the function $u\left( x,t\right) $
with respect to $t$ and do not change notations for brevity, $u\left(
x,-t\right) :=u\left( x,t\right) ,t\in \left( 0,T\right) .$ Let $Q_{T}^{\pm
}=\Omega \times \left( -T,T\right) ,S_{T}^{\pm }=\partial \Omega \times
\left( -T,T\right) .$ Obviously $\left\Vert u\mid _{S_{T}^{\pm }}\right\Vert
_{H^{1}\left( S_{T}^{\pm }\right) }=2\left\Vert u\mid _{S_{T}}\right\Vert
_{H^{1}\left( S_{T}\right) }$ and $\left\Vert \partial _{\nu }u\mid
_{S_{T}^{\pm }}\right\Vert _{L_{2}\left( S_{T}^{\pm }\right) }=2\left\Vert
\partial _{\nu }u\mid _{S_{T}}\right\Vert _{L_{2}\left( S_{T}\right) },$
where $\partial _{\nu }$ means the normal derivative. The Lipschitz
stability estimate for the Cauchy problem with the lateral data is%
\begin{equation}
\left\Vert u\right\Vert _{H^{1}\left( Q_{T}^{\pm }\right) }\leq C\left[
\left\Vert u\mid _{S_{T}}\right\Vert _{H^{1}\left( S_{T}\right) }+\left\Vert
\partial _{\nu }u\mid _{S_{T}}\right\Vert _{L_{2}\left( S_{T}\right) }\right]
\label{2.0}
\end{equation}%
with a certain constant $C>0$ independent on the function $u$. Hence, the
trace theorem implies the Lipschitz stability estimate for the function $f$
with a different constant $C,$%
\begin{equation*}
\left\Vert u\left( x,0\right) \right\Vert _{L_{2}\left( \Omega \right)
}=\left\Vert f\right\Vert _{L_{2}\left( \Omega \right) }\leq C\left[
\left\Vert u\mid _{S_{T}}\right\Vert _{H^{1}\left( S_{T}\right) }+\left\Vert
\partial _{\nu }u\mid _{S_{T}}\right\Vert _{L_{2}\left( S_{T}\right) }\right]
.
\end{equation*}

Estimate (\ref{2.0}) plays a fundamental role in the control theory, since
it is used for proofs of exact controllability theorems. For the first time
estimate (\ref{2.0}) was proved in 1986 in \cite{L} for equation (\ref{1.0})
with $c\equiv 1$ with the aim of applying to the control theory. However,
the method of multipliers, which was proposed in \cite{L}, cannot handle
neither variable lower order terms of the operator $L$ nor a variable
coefficient $c\left( x\right) .$ On the other hand, Carleman estimates are
not sensitive to lower order terms of PDE operators and also can handle the
case of a variable coefficient $c\left( x\right) .$

In \cite{KlibM} the Carleman estimate was applied for the first time to
obtain (\ref{2.0}). In \cite{KlibM} (\ref{2.0}) was proved for the case of
the hyperbolic equation (\ref{1.6}) with $L=\Delta +$ (\emph{variable lower
order terms}). Next, the result of \cite{KlibM} was extended in \cite{Kaz,K}
to a more general case of the hyperbolic inequality%
\begin{equation}
\left\vert u_{tt}-\Delta u\right\vert \leq A\left[ \left\vert \nabla
u\right\vert +\left\vert u_{t}\right\vert +\left\vert u\right\vert
+\left\vert f\right\vert \right] \text{ in }Q_{T},  \label{2.1}
\end{equation}%
where $A=const.>0$ and $f\in L_{2}\left( Q_{T}\right) .$ Although in
publications \cite{Kaz,KlibM,K} $c\equiv 1,$ it is clear from them that the
key idea is in applying the Carleman estimate, while a specific form of the
principal part of the hyperbolic operator is less important. This thought is
reflected in the proof of Theorem 3.4.8 of the book \cite{Is}. Thus, the
Lipschitz stability estimate (\ref{2.0}) for the variable coefficient $%
c\left( x\right) $ was obtained in section 2.4 of the book \cite{KT} as well
as in \cite{ClK}. In particular, in \cite{KT} the case of an analog of the
hyperbolic inequality (\ref{2.1}) was considered, where $\left\vert
u_{tt}-\Delta u\right\vert $ was replaced with $\left\vert c^{-2}\left(
x\right) u_{tt}-\Delta u\right\vert $. The idea of \cite{Kaz} was used in
the control theory in, e.g. \cite{LT3,LT4}.

In the case of parabolic and elliptic operators, Carleman estimates are
known for rather arbitrary variable principal parts \cite{Is,KT,LRS}. On the
other hand, it is well known that in the hyperbolic case the Carleman
estimate can be effectively analytically verified for a generic operator $%
\partial _{t}^{2}-L_{0}$ only if $L_{0}=c^{2}\left( x\right) \Delta ,$ and a
condition like
\begin{equation}
\left( x-x_{0},\nabla \left( c^{-2}\left( x\right) \right) \right) \geq
0,\forall x\in \overline{\Omega }\text{ }  \label{1.30}
\end{equation}%
holds. In (\ref{1.30}) $x_{0}$ is a certain point and $\left( ,\right) $ is
the scalar product in $\mathbb{R}^{n}.$ This is the reason why the above
mentioned Lipschitz stability estimates were established only using
assumptions like the one in (\ref{1.30}). Clearly, (\ref{1.30}) holds for $%
c\equiv const.\neq 0$. See, e.g. Theorem 1.10.2 in \cite{BKbook} for the
proof of the Carleman estimate with condition (\ref{1.30}). A more general
case of condition (\ref{1.30}) can be found in Theorem 3.4.1 of \cite{Is}.
The second way of proving Lipschitz stability estimates is via imposing some
conditions of the Riemannian geometry on coefficients of the operator $L_{0}$
\cite{Bardos,LT2,Rom1,Rom2,SU}. Publications \cite{LT2,Rom1,Rom2} use some
modifications of the idea of \cite{Kaz,KlibM}; in particular, the case of
hyperbolic inequality was considered in \cite{Rom1}. Unlike (\ref{1.30}),
conditions of the Riemannian geometry cannot be effectively analytically
verified for an operator $L_{0}$ with generic coefficients, e.g. $%
L_{0}=c^{2}\left( x\right) \Delta .$ A slight variation of (\ref{1.30})
guarantees the non-trapping condition, see formula (3.24) in \cite{Rom}.
Uniqueness theorems for TAT were also obtained in \cite{AK,FR,SU} for the
case (\ref{1.0}), (\ref{1.01}).

In addition, to the Lipschitz stability, the Quasi-Reversibility Method
(QRM) for the above mentioned Cauchy problem with the lateral data was
developed in \cite{KlibM} and numerically tested in \cite{ClK,KR,KKKN}. We
refer to \cite{LL} for the originating work on QRM. The convergence of the
QRM solution to the exact solution was proven on the basis of the above
Lipschitz stability results. Numerical testing has consistently demonstrated
a high degree of robustness. In particular, accurate results were obtained
in \cite{KKKN} with up to 50\% noise in the data. Some other numerical
methods were proposed in \cite{AK,SU}. Convergence of all numerical methods
mentioned in this paragraph was proven only for the complete data collection
case of IP1 with $L_{0}=c^{2}\left( x\right) \Delta $ and under some
restrictive conditions imposed on the function $c\left( x\right) $.

\section{Logarithmic Stability}

\label{sec:2}

\subsection{Transformation}

\label{sec:2.1}

First, we consider the following well known Laplace-like transformation \cite%
{KT,LRS}, which transforms the hyperbolic Cauchy problem in a similar
parabolic Cauchy problem,
\begin{equation}
\mathcal{L}g=\overline{g}\left( t\right) =\frac{1}{\sqrt{\pi t}}%
\int\limits_{0}^{\infty }\exp \left( -\frac{\tau ^{2}}{4t}\right) g\left(
\tau \right) d\tau .  \label{1.10}
\end{equation}%
The transformation (\ref{1.10}) is an analog of the Laplace transform, and
it is one-to-one. It is valid for, e.g. all functions $g\in C\left[ 0,\infty
\right) $ which satisfy $\left\vert g\left( t\right) \right\vert \leq
A_{g}e^{k_{g}t},$ where $A_{g}$ and $k_{g}$ are positive constants depending
on $g$. It follows from (\ref{1.171})\ that the solution $u\left( x,t\right)
$ of the problem (\ref{1.6}), (\ref{1.7}) satisfies this condition together
with its derivatives up to the third order. Obviously%
\begin{equation*}
\frac{\partial }{\partial t}\left[ \frac{1}{\sqrt{\pi t}}\exp \left( -\frac{%
\tau ^{2}}{4t}\right) \right] =\frac{\partial ^{2}}{\partial \tau ^{2}}\left[
\frac{1}{\sqrt{\pi t}}\exp \left( -\frac{\tau ^{2}}{4t}\right) \right] .
\end{equation*}%
Hence,%
\begin{equation}
\mathcal{L}\left( g^{\prime \prime }\right) =\overline{g}^{\prime }\left(
t\right) ,\forall g\in C^{2}\left[ 0,\infty \right) \text{ such that }%
g^{\prime }\left( 0\right) =0.  \label{1.101}
\end{equation}%
Changing variables in (\ref{1.10}) $\tau \Leftrightarrow z,\tau /2\sqrt{t}%
:=z,$ we obtain $\lim_{t\rightarrow 0^{+}}\overline{g}\left( t\right)
=g\left( 0\right) .$ Denote $v:=\mathcal{L}u.$ It follows from (\ref{1.171})
and (\ref{1.101}) that
\begin{equation}
v\in C^{2+\alpha ,1+\alpha /2}\left( \mathbb{R}^{n}\times \left[ 0,T\right]
\right) ,\forall \alpha \in \left( 0,1\right) ,\forall T>0.  \label{1.102}
\end{equation}%
By (\ref{1.6}), (\ref{1.7}) and (\ref{1.102}) the function $v\left(
x,t\right) $ is the solution of the following parabolic Cauchy problem%
\begin{eqnarray}
v_{t} &=&Lv,x\in \mathbb{R}^{n},t>0,  \label{1.11} \\
v\left( x,0\right) &=&f\left( x\right) .  \label{1.12}
\end{eqnarray}%
We refer here to the well known uniqueness result for the solution $v\in
C^{2+\alpha ,1+\alpha /2}\left( \mathbb{R}^{n}\times \left[ 0,T\right]
\right) ,\forall T>0$ of the problem (\ref{1.11}), (\ref{1.12}) \cite{LSU}.

Below we work only with the function $v$. As to this function, we set
everywhere below $T:=1$ for the sake of definiteness. Denote%
\begin{equation}
\mathcal{L\varphi }_{1}=\overline{\varphi }_{1}\left( x,t\right) =v\mid
_{S_{1}},\text{ }\mathcal{L\varphi }_{2}=\overline{\varphi }_{2}\left(
x,t\right) =v\mid _{P_{1}}.  \label{1.13}
\end{equation}%
Then%
\begin{equation}
\overline{\varphi }_{1}\in C^{2+\alpha ,1+\alpha /2}\left( \overline{S}%
_{1}\right) ,\overline{\varphi }_{2}\in C^{2+\alpha ,1+\alpha /2}\left(
\overline{P}_{1}\right) .  \label{1.14}
\end{equation}%
Let
\begin{equation}
\overline{\psi }_{1}\left( x,t\right) =\partial _{\nu }v\mid _{S_{1}},%
\overline{\psi }_{2}\left( x,t\right) =\partial _{x_{1}}v\mid _{P_{1}}.
\label{1.140}
\end{equation}%
By Theorem 5.2 of Chapter IV of \cite{LSU}, (\ref{1.171}) and (\ref{1.13})-(%
\ref{1.140}) there exist numbers $C\left( \Omega ,L\right) ,C\left(
P,L\right) >0$ depending only on listed parameters such that%
\begin{eqnarray}
\left\Vert \overline{\psi }_{1}\right\Vert _{C^{1+\alpha ,\alpha /2}\left(
\overline{S}_{1}\right) } &\leq &C\left( \Omega ,L\right) \left\Vert
\overline{\varphi }_{1}\right\Vert _{C^{2+\alpha ,1+\alpha /2}\left(
\overline{S}_{1}\right) },  \label{1.14_1} \\
\left\Vert \overline{\psi }_{2}\right\Vert _{C^{1+\alpha ,\alpha /2}\left(
\overline{P}_{1}\right) } &\leq &C\left( P,L\right) \left\Vert \overline{%
\varphi }_{2}\right\Vert _{C^{2+\alpha ,1+\alpha /2}\left( \overline{P}%
_{1}\right) }.  \label{1.141}
\end{eqnarray}

We now describe an elementary and well known procedure of finding the normal
derivative of the function $v$ either at $S_{1}$ (in the case of IP1) or at $%
P_{1}$ (in the case of IP2). In the case of IP1 we solve the initial
boundary value problem for equation (\ref{1.11}) for $\left( x,t\right) \in
\left( \mathbb{R}^{n}\diagdown \Omega \right) \times \left( 0,1\right) $
with the zero initial condition in $\mathbb{R}^{n}\diagdown \Omega $
(because of (\ref{1.5})) and the Dirichlet boundary condition $v\mid
_{S_{1}}=\overline{\varphi }_{1}.$ Then we uniquely find the normal
derivative $\partial _{\nu }v\mid _{S_{1}}=\overline{\psi }_{1}$. Similarly,
in the case of IP2, we uniquely find the Neumann boundary condition $%
\partial _{x_{1}}v\mid _{P_{1}}=\overline{\psi }_{2}$. Estimates (\ref%
{1.14_1}), (\ref{1.141}) ensure the stability of this procedure.

Therefore, each problem IP1, IP2 is replaced with a problem for the
parabolic PDE (\ref{1.11}) with the lateral Cauchy data. These data are
given at $S_{1}$ for IP1 and at $P_{1}$ for IP2. Uniqueness of the solution
of each of these parabolic inverse problems follows from standard theorems
about uniqueness of the continuation of solutions of parabolic PDEs with the
data at the lateral surface \cite{Is,KT,LRS}.

In stability estimates one is usually interested to see how the solution
varies for a small variation of the input data. Therefore, following (\ref%
{1.171}), (\ref{1.8}) and (\ref{1.9}), we assume that in the case of IP1%
\begin{equation}
\left\Vert \varphi _{1}\right\Vert _{C^{3}\left( \overline{S}_{T}\right)
}\leq \delta e^{dT},\forall T>0,  \label{1.142}
\end{equation}%
and in the case of IP2%
\begin{equation}
\left\Vert \varphi _{2}\right\Vert _{C^{3}\left( \overline{P}_{T}\right)
}\leq \delta e^{dT},\forall T>0,  \label{1.143}
\end{equation}%
where $\delta \in \left( 0,1\right) $ is a sufficiently small number. Note
that it is not necessary that $\delta =B,$ where $B$ is the number from (\ref%
{1.171}). Indeed, while the number $B$ in (\ref{1.171}) is not assumed to be
sufficiently small and is involved in the estimate of the norm $\left\Vert
u\right\Vert _{C^{3}\left( \mathbb{R}^{n}\times \left[ 0,T\right] \right)
},\forall T>0$ in the entire space, the number $\delta $ is a part of the
estimate of the norm of the boundary data for either of above inverse
problems. Using (\ref{1.10}), (\ref{1.101}) and (\ref{1.13})-(\ref{1.143}),
we obtain%
\begin{eqnarray}
\left\Vert \overline{\varphi }_{1}\right\Vert _{C^{2+\alpha ,1+\alpha
/2}\left( \overline{S}_{1}\right) }+\left\Vert \overline{\psi }%
_{1}\right\Vert _{C^{1+\alpha ,\alpha /2}\left( \overline{S}_{1}\right) }
&\leq &C\left( \Omega ,L,d\right) \delta ,  \label{1.144} \\
\left\Vert \overline{\varphi }_{2}\right\Vert _{C^{2+\alpha ,1+\alpha
/2}\left( \overline{P}_{1}\right) }+\left\Vert \overline{\psi }%
_{2}\right\Vert _{C^{1+\alpha ,\alpha /2}\left( \overline{P}_{1}\right) }
&\leq &C\left( P,L,d\right) \delta ,  \label{1.145}
\end{eqnarray}%
where constants $C\left( \Omega ,L,d\right) ,C\left( P,L,d\right) >0$ depend
only on listed parameters. It follows from (\ref{1.144}) that with a
different constant $\overline{C}:=\overline{C}\left( \Omega ,L,d\right) >0$%
\begin{equation}
\left\Vert \overline{\varphi }_{1}\right\Vert _{H^{1}\left( S_{1}\right)
}+\left\Vert \overline{\psi }_{1}\right\Vert _{L_{2}\left( S_{1}\right)
}\leq \overline{C}\delta .  \label{1.146}
\end{equation}

\textbf{Remarks 2.1}.

\textbf{1}. The number $\delta $ can be viewed as an upper estimate of the
level of error in the data $\varphi _{1},\varphi _{2}.$ Hence, Theorems 2.1,
2.2 below address the question of estimating variations of the solution $f$
of either IP1 or IP2 via the upper estimate of the level of error in the
data.

\textbf{2}. Since the kernel of the transform $\mathcal{L}$ decays rapidly
with $\tau \rightarrow \infty ,$ then the condition $t\in \left( 0,\infty
\right) $ in (\ref{1.8}), (\ref{1.9}) is not a serious restriction from the
applied standpoint. In addition, if having the data in (\ref{1.8}), (\ref%
{1.9}) only on a finite time interval $t\in \left( 0,T\right) $ and knowing
an upper estimate of a norm of the function $f$ in (\ref{1.7}), one can
estimate the error in the integral (\ref{1.10}) when integrating over $\tau
\in \left( T,\infty \right) .$ This error will be small if either $T$ is
large or $t$ is small in (\ref{1.10}), (\ref{1.11}). Next, this error can be
incorporated in the stability estimates of theorems of this section.

\textbf{3}. Another argument about $t\in \left( 0,\infty \right) $ comes
from the recent experience of the author of working with time resolved real
data for wave processes \cite{BKbook}. The author has learned that almost
all time resolved experimental data for wave processes in non-attenuating
media are highly oscillatory due to some unknown processes in measurement
devices, see graphs of those data in these references. Because of high
oscillations, these data are not governed by a hyperbolic PDE even for the
case of the free space, where the wave equation is supposed to work (see the
graphs of experimental data in chapters 5 and 6 of \cite{BKbook}).
Therefore, the first step to make the inverse algorithm work was to
preprocess the experimental data via a new data preprocessing procedure.
This procedure uses only a small portion of the real data and immerses it in
a specially processed data for the uniform medium. Since the case of the
uniform medium can be solved analytically, then there is no problem to know
the immersed data for all $t\in \left( 0,\infty \right) .$ Since accurate
imaging results were obtained in \cite{BKbook} for the case of blind
experimental data, then that data preprocessing procedure was unbiased.

\subsection{Logarithmic stability estimate for Inverse Problem 1}

\label{sec:2.2}

To prove convergence of the QRM (Theorem 3.1), it is convenient to consider
a parabolic inequality in the integral form, which is more general than
equation (\ref{1.11}). Consider the function $w\in C^{2,1}\left( \overline{Q}%
_{1}\right) $ satisfying the following inequality
\begin{equation}
\int\limits_{Q_{1}}\left( w_{t}-Lw\right) ^{2}dxdt\leq K^{2},K=const.\geq 0.
\label{1.15}
\end{equation}

\textbf{Theorem 2.1.} \emph{Let conditions (\ref{1.2})-(\ref{1.4}) be
fulfilled. Let the function }$w\in C^{2,1}\left( \overline{Q}_{1}\right) $%
\emph{\ satisfies inequality (\ref{1.15}). Denote }%
\begin{eqnarray}
g\left( x\right) &=&w\left( x,0\right) ,\beta _{0}\left( x,t\right) =w\mid
_{S_{1}},\beta _{1}\left( x,t\right) =\partial _{\nu }w\mid _{S_{1}},  \notag
\\
F &=&\left\Vert \beta _{0}\right\Vert _{H^{1}\left( S_{1}\right)
}+\left\Vert \beta _{1}\right\Vert _{L_{2}\left( S_{1}\right) }+K.
\label{1.19}
\end{eqnarray}%
\emph{\ \ Assume that an upper bound }$C_{1}=const.>0$ \emph{for the norm }$%
\left\Vert \nabla g\right\Vert _{L_{2}\left( \Omega \right) }$ \emph{is
known,}%
\begin{equation}
\left( \sum\limits_{i=1}^{n}\left\Vert g_{x_{i}}\right\Vert _{L_{2}\left(
\Omega \right) }^{2}\right) ^{1/2}:=\left\Vert \nabla g\right\Vert
_{L_{2}\left( \Omega \right) }\leq C_{1}.  \label{1.18}
\end{equation}%
\emph{Then there exist a constant }$M=M\left( L,\Omega \right) >0$\emph{\
and a sufficiently small number }$\delta _{0}=\delta _{0}\left( L,\Omega
,C_{1}\right) \in \left( 0,1\right) ,$ \emph{both dependent only on listed
parameters,\ such that\ if }$F\in \left( 0,\delta _{0}\right) $, \emph{then}
\emph{the following logarithmic stability estimate is valid }%
\begin{equation}
\left\Vert g\right\Vert _{L_{2}\left( \Omega \right) }\leq \frac{MC_{1}}{%
\sqrt{\ln \left( F^{-1}\right) }}.  \label{1.20}
\end{equation}%
\emph{In particular, in the case of IP1, let Assumption holds and (\ref{1.5}%
), (\ref{1.142}) be valid. Suppose that the number }$\delta $ \emph{in (\ref%
{1.142}) is so small that }$\overline{C}\delta \in \left( 0,\delta
_{0}\right) $, \emph{where} $\overline{C}=\overline{C}\left( \Omega
,L,d\right) >0$ \emph{is the number in (\ref{1.146}). Also, assume that the
upper bound }$C_{1}$\emph{\ of the norm }$\left\Vert \nabla f\right\Vert
_{L_{2}\left( \Omega \right) }$\emph{\ is given, \ }%
\begin{equation}
\left\Vert \nabla f\right\Vert _{L_{2}\left( \Omega \right) }\leq C_{1}.
\label{1.21}
\end{equation}%
\emph{Then}%
\begin{equation}
\left\Vert f\right\Vert _{L_{2}\left( \Omega \right) }\leq \frac{MC_{1}}{%
\sqrt{\ln \left[ \left( \overline{C}\delta \right) ^{-1}\right] }}.\text{ }
\label{1.22}
\end{equation}

\textbf{Proof}. In this proof\textbf{\ }$M=M\left( L,\Omega \right) >0$
denotes a generic positive constant depending only on $L,\Omega .$ First, we
prove (\ref{1.20}). Let $\varkappa \in \left( 0,1\right) $ be an arbitrary
number. Then it follows from Theorem 2 of \cite{Kl1} that there exists a
constant $r=r\left( L,\Omega ,\varkappa \right) \in \left( 0,1\right) $ such
that%
\begin{equation}
\left\Vert g\right\Vert _{L_{2}\left( \Omega \right) }\leq \frac{MC_{1}}{%
\varkappa \sqrt{\ln \left[ \left( rF\right) ^{-1}\right] }}+M\left( \frac{1}{%
r}\right) ^{\varkappa }F^{1-\varkappa },  \label{3.1}
\end{equation}%
as long as $F\in \left( 0,1\right) .$ We can fix $\varkappa $ via, e.g.
setting $\varkappa :=1/2.$ It is clear therefore that there exists a
sufficiently small number $\delta _{0}=\delta _{0}\left( L,\Omega
,C_{1}\right) \in \left( 0,1\right) $ such that if $F\in \left( 0,\delta
_{0}\right) ,$ then (\ref{3.1}) implies (\ref{1.20}).

We now prove (\ref{1.22}). It follows from (\ref{1.11}) that (\ref{1.15})
holds for the function $w:=v$ with $K=0.$ As it was shown above, (\ref{1.146}%
) follows from (\ref{1.142}). Hence, using (\ref{1.146}) and (\ref{1.19}),
we obtain $F\leq \overline{C}\delta .$ Hence, (\ref{1.22}) follows from (\ref%
{1.20}). $\square $

\textbf{Remark 2.2. }Estimates (\ref{1.20}), (\ref{1.22}) are the so-called
\textquotedblleft conditional stability estimates", which is often the case
in ill-posed problems \cite{BKbook,LRS,T}. For another example we refer to H%
\"{o}lder stability estimates for solutions of some ill-posed problems for
PDEs, see, e.g. \cite{Is,KT,LRS}.\ The knowledge of the upper bound $C_{1}$
for the gradient in (\ref{1.18}), (\ref{1.21}) corresponds well with the
Tikhonov concept of compact sets as sets of \textquotedblleft admissible"
solutions of ill-posed problems \cite{Bak,BKbook,EHN,LRS,T}. Indeed, since
by (\ref{1.5}) $f\mid _{\partial \Omega }=0,$ then $\left\Vert f\right\Vert
_{L_{2}\left( \Omega \right) }\leq \overline{R}\left\Vert \nabla
f\right\Vert _{L_{2}\left( \Omega \right) }\leq \overline{R}C_{1},$ where
the constant $\overline{R}>0$ depends only on the domain $\Omega .$ Thus, in
this case the function $f$ \ belongs to a compact set in $L_{2}\left( \Omega
\right) ,$ and this set is determined by the constant $C_{1}.$

\subsection{Logarithmic stability estimate for Inverse Problem 2}

\label{sec:2.3}

The logarithmic stability estimate of the paper \cite{Kl2} in the infinite
domain was obtained for the case of the pointwise inequality
\begin{equation}
\left\vert v_{t}-L_{0}v\right\vert \leq A\left( \left\vert \nabla
v\right\vert +\left\vert v\right\vert \right) ,A=const.>0,  \label{3.1_0}
\end{equation}%
where the operator $L_{0}$ is defined in (\ref{1.2_1}). However, to prove
convergence of the numerical method of section 3, we need to estimate the
initial condition for the case of the integral inequality, like the one in (%
\ref{1.15}). The Carleman estimate of \cite{Kl2} is not a standard one.
Indeed, unlike the standard Carleman estimate for the parabolic operator
\cite{BKbook,KT,LRS}, the integration domain of \cite{Kl2} is a part of the
strip $\left\{ \left\vert t-\varepsilon \right\vert <\tau \varepsilon ,\tau
\in \left( 0,1\right) \right\} ,$ and that Carleman estimate does not break
when $\varepsilon \rightarrow 0^{+}.$

There \ are two main differences between Theorem 2.2 (below) and Theorem 1
of \cite{Kl2}. First, we work now with the integral inequality instead of
the pointwise inequality (\ref{3.1_0}) of \cite{Kl2}. Second, it is assumed
in \cite{Kl2} that the inequality (\ref{3.1_0}) is valid in $\Theta \times
\left( 0,T\right) ,$ where $\Theta \subseteq \mathbb{R}^{n}$ is an unbounded
domain. It is also assumed that the Dirichlet boundary condition $v\mid
_{\partial \Theta \times \left( 0,T\right) }=0.$ Unlike this, Theorem 2.2
does not use the assumption about the knowledge of this Dirichlet boundary
condition.

Denote $\overline{x}=\left( x_{2},...,x_{n}\right) .$ Below we specify
numbers $1/4,1/2,3/4$ for brevity only.\ In fact, some other numbers,
respectively $\eta _{1}<\eta _{2}<\eta _{3}<1$ from the interval $\left(
0,1\right) $ can be used. Changing variables $\left( x^{\prime },t^{\prime
}\right) =\left( \sqrt{c}x,dt\right) $ with an appropriate constant $c>0$
and keeping the same notations for new variables for brevity, we obtain that
\begin{equation}
\Omega \subset \left\{ x_{1}+\left\vert \overline{x}\right\vert ^{2}<\frac{1%
}{4},x_{1}>0\right\} .  \label{3.2}
\end{equation}%
Let $\varepsilon \in \left( 0,1\right) $ be a sufficiently small number.
Consider the following functions $\psi \left( x,t\right) ,\varphi \left(
x,t\right) ,$%
\begin{eqnarray}
\psi \left( x,t\right) &=&x_{1}+\left\vert \overline{x}\right\vert ^{2}+%
\frac{\left( t-\varepsilon \right) ^{2}}{\varepsilon ^{2}}+\frac{1}{4},
\label{3.3} \\
\varphi \left( x,t\right) &=&\exp \left( \frac{\psi ^{-\nu }}{\varepsilon }%
\right) ,  \label{3.4}
\end{eqnarray}%
where $\nu >1$ is a large parameter which will be defined later. The
function $\varphi \left( x,t\right) $ is the Carleman Weight Function (CWF)
in the Carleman estimate of Lemma 3.1. The main difference between $\varphi
\left( x,t\right) $ in (\ref{3.4}) and the standard CWF for the parabolic
operator \cite{BKbook,KT,LRS} is that the small parameter $\varepsilon $ is
involved in both functions $\psi \left( x,t\right) $ and $\varphi \left(
x,t\right) $. Denote%
\begin{eqnarray}
G_{3/4} &=&\left\{ \left( x,t\right) :\psi \left( x,t\right) <\frac{3}{4}%
,x_{1}>0\right\} ,  \label{3.5} \\
G_{1/2} &=&\left\{ \left( x,t\right) :\psi \left( x,t\right) <\frac{1}{2}%
,x_{1}>0\right\} .  \label{3.6}
\end{eqnarray}%
Using (\ref{3.2})-(\ref{3.6}), we obtain%
\begin{eqnarray}
G_{1/2} &\subset &G_{3/4},\varphi ^{2}\left( x,t\right) \geq \exp \left[
\frac{2^{\nu +1}}{\varepsilon }\right] \text{ in }G_{1/2},  \label{3.7} \\
G_{3/4} &\subset &\left\{ \left\vert t-\varepsilon \right\vert <\frac{%
\varepsilon }{\sqrt{2}}\right\} \subset \left\{ t\in \left( 0,1\right)
\right\} ,  \label{3.71} \\
\Omega &\subset &RG_{1/2}\subset RG_{3/4},  \label{3.8} \\
\partial G_{3/4} &=&\partial _{1}G_{3/4}\cup \partial _{2}G_{3/4},\partial
_{1}G_{3/4}=\left\{ x_{1}=0\right\} \cap \overline{G}_{3/4},\partial
_{2}G_{3/4}=\left\{ \psi \left( x,t\right) =\frac{3}{4},x_{1}>0\right\} .
\label{3.9}
\end{eqnarray}%
In (\ref{3.8}) $RG_{1/2}$ and $RG_{3/4}$ are orthogonal projections of
domain $G_{1/2}$ and $G_{3/4}$ respectively on the hyperplane $\left\{
t=0\right\} .$ The same notation $RH$ is kept below for the projection of
any other domain $H\subset \left[ \mathbb{R}^{n}\times \left( 0,1\right) %
\right] $ on the hyperplane $\left\{ t=0\right\} .$ Denote%
\begin{eqnarray}
\Phi &=&\left\{ \left( x,t\right) :x_{1}\in \left( 0,1\right) ,\left(
x_{2},x_{3},...,x_{n}\right) \in \left( -1,1\right) ^{n-1},t\in \left(
0,1\right) \right\} ,  \label{3.91} \\
\partial _{1}\Phi &=&\overline{\Phi }\cap P=\left\{ \left( x,t\right)
:x_{1}=0,\overline{x}\in \left( -1,1\right) ^{n-1},t\in \left( 0,1\right)
\right\} .  \label{3.92}
\end{eqnarray}%
By (\ref{3.5}), (\ref{3.71}) and (\ref{3.9})-(\ref{3.92})
\begin{equation}
\partial _{1}G_{3/4}\subset \partial _{1}\Phi .  \label{3.93}
\end{equation}%
Recall that (\ref{1.143}) implies (\ref{1.145}). Hence, assuming that (\ref%
{1.143}) holds and using (\ref{3.92}), we derive, similarly with the above
derivation of (\ref{1.146}) from (\ref{1.142}), (\ref{1.144}), that there
exists a constant $\widetilde{C}=\widetilde{C}\left( P,L,\Phi ,b\right) >0$
such that
\begin{equation}
\left\Vert \overline{\varphi }_{2}\right\Vert _{H^{1}\left( \partial
_{1}\Phi \right) }+\left\Vert \overline{\psi }_{2}\right\Vert _{L_{2}\left(
\partial _{1}\Phi \right) }\leq \widetilde{C}\delta .  \label{3.94}
\end{equation}

Everywhere below $C=C\left( L_{0},RG_{3/4}\right) >0$ and $M_{1}=M_{1}\left(
L,\Phi \right) >0$\emph{\ }denote different positive constants depending
only on listed parameters. The following lemma is follows immediately from
Theorem 2 of \cite{Kl2} and (\ref{3.9}).

\textbf{Lemma 2.1}.\emph{\ Let coefficients of the operator }$L_{0}$\emph{\
in (\ref{1.2_1}) satisfy conditions (\ref{1.3}), (\ref{1.4}). Then there
exist a sufficiently large constant }$\nu _{0}=\nu _{0}\left(
L_{0},RG_{3/4}\right) >1$\emph{\ and a sufficiently small number }$%
\varepsilon _{0}=\varepsilon _{0}\left( L_{0},RG_{3/4}\right) \in \left(
0,1\right) ,$ \emph{both dependent only on }$L_{0}$\emph{\ and} $PG_{3/4},$%
\emph{\ such that the following Carleman estimate holds }
\begin{eqnarray*}
&&\frac{C\nu ^{3}}{\varepsilon ^{3}}\exp \left( \frac{2\cdot 4^{\nu }}{%
\varepsilon }\right) \int\limits_{\partial _{1}G_{3/4}}\left(
u^{2}+\left\vert \nabla u\right\vert ^{2}+u_{t}^{2}\right) d\overline{x}dt \\
&&+\frac{C\nu ^{3}}{\varepsilon ^{3}}\left( \frac{4}{3}\right) ^{2\nu }\exp %
\left[ \frac{2}{\varepsilon }\left( \frac{4}{3}\right) ^{\nu }\right]
\int\limits_{\partial _{2}G_{3/4}}\left( u^{2}+\left\vert \nabla
u\right\vert ^{2}+u_{t}^{2}\right) d\sigma +\int\limits_{G_{3/4}}\left(
u_{t}-L_{0}u\right) ^{2}\varphi ^{2}\left( x,t\right) dxdt \\
&\geq &C\int\limits_{G_{3/4}}\left( \frac{\nu }{\varepsilon }\left\vert
\nabla u\right\vert ^{2}+\frac{\nu ^{4}}{\varepsilon ^{3}}\psi ^{-2\nu
}u^{2}\right) \varphi ^{2}\left( x,t\right) dxdt,\forall \nu \geq \nu
_{0},\forall \varepsilon \in \left( 0,\varepsilon _{0}\right) ,\forall u\in
C^{2,1}\left( \overline{G}_{3/4}\right) .
\end{eqnarray*}

By (\ref{3.71}) Lemma 2.1 provides the Carleman estimate in the narrow strip
$\left\{ \left\vert t-\varepsilon \right\vert <\varepsilon /\sqrt{2}\right\}
.$ At the same time, it is also important in numerical studies of the QRM to
estimate its solution in a not narrow strip. This can be done via the
standard Carleman estimate. Therefore, we introduce now notations, which are
similar with (\ref{3.3})-(\ref{3.9}), except that a narrow strip with
respect to $t$ is not used. Let
\begin{eqnarray}
\theta \left( x,t\right) &=&x_{1}+\left\vert \overline{x}\right\vert
^{2}+\left( t-\frac{1}{2}\right) ^{2}+\frac{1}{4},  \label{3.100} \\
\xi \left( x,t\right) &=&\exp \left( \lambda \theta ^{-\nu }\right) ,
\label{3.101}
\end{eqnarray}%
where $\lambda >1$ is a large parameter which is chosen later. Denote%
\begin{eqnarray}
D_{3/4} &=&\left\{ \left( x,t\right) :\theta \left( x,t\right) <\frac{3}{4}%
,x_{1}>0\right\} ,  \label{3.102} \\
D_{1/2} &=&\left\{ \left( x,t\right) :\theta \left( x,t\right) <\frac{1}{2}%
,x_{1}>0\right\} ,  \label{3.103} \\
\partial D_{3/4} &=&\partial _{1}D_{3/4}\cup \partial _{2}D_{3/4},\partial
_{1}D_{3/4}=\left\{ x_{1}=0\right\} \cap \overline{D}_{3/4},\partial
_{2}D_{3/4}=\left\{ \psi \left( x,t\right) =\frac{3}{4},x_{1}>0\right\} .
\label{3.104}
\end{eqnarray}%
Using (\ref{3.2}), (\ref{3.8}), (\ref{3.92}), (\ref{3.100}) and (\ref{3.102}%
)-(\ref{3.104}), we obtain%
\begin{eqnarray}
\Omega &\subset &RD_{3/4}=RG_{3/4},  \label{3.1031} \\
D_{3/4} &\subset &\left\{ \left\vert t-\frac{1}{2}\right\vert <\frac{1}{2}%
\right\} \subset \left\{ t\in \left( 0,1\right) \right\} ,  \label{3.1032} \\
D_{3/4} &\subset &\Phi .  \label{3.1033}
\end{eqnarray}%
Lemma 2.2 follows from the Carleman estimate for the parabolic operator of
Lemma 3 of \S 1 of Chapter 4 of the book \cite{LRS} as well as from (\ref%
{3.1031}).

\textbf{Lemma 2.2}. \emph{\ Let coefficients of the operator }$L_{0}$\emph{\
in (\ref{1.2_1}) satisfy conditions (\ref{1.3}), (\ref{1.4}). Then there
exist sufficiently large constants }$\nu _{0}=\nu _{0}\left(
L_{0},RG_{3/4}\right) >1,\lambda _{0}=\lambda _{0}\left(
L_{0},RG_{3/4}\right) >1$\emph{\ such that the following Carleman estimate
holds }
\begin{eqnarray*}
&&C\lambda ^{3}\nu ^{3}\exp \left( 2\lambda \cdot 4^{\nu }\right)
\int\limits_{\partial _{1}D_{3/4}}\left( u^{2}+\left\vert \nabla
u\right\vert ^{2}+u_{t}^{2}\right) d\overline{x}dt \\
&&+C\lambda ^{3}\nu ^{3}\left( \frac{4}{3}\right) ^{2\nu }\exp \left[
2\lambda \left( \frac{4}{3}\right) ^{\nu }\right] \int\limits_{\partial
_{2}D_{3/4}}\left( u^{2}+\left\vert \nabla u\right\vert
^{2}+u_{t}^{2}\right) d\sigma +\int\limits_{D_{3/4}}\left(
u_{t}-L_{0}u\right) ^{2}\xi ^{2}\left( x,t\right) dxdt \\
&\geq &C\int\limits_{D_{3/4}}\left( \lambda \nu \left\vert \nabla
u\right\vert ^{2}+\lambda ^{3}\nu ^{4}\psi ^{-2\nu }u^{2}\right) \xi
^{2}\left( x,t\right) dxdt,\forall \nu \geq \nu _{0},\forall \lambda \geq
\lambda _{0},\forall u\in C^{2,1}\left( \overline{D}_{3/4}\right) .
\end{eqnarray*}

\textbf{Theorem 2.2.} \emph{Let conditions (\ref{1.2})-(\ref{1.4}) and (\ref%
{3.2}) be valid. Suppose that the function }$w\in C^{2,1}\left( \overline{Q}%
\right) $\emph{\ satisfies the following integral inequality\ }
\begin{equation}
\int\limits_{Q}\left( w_{t}-Lw\right) ^{2}dxdt\leq K^{2},K=const.\geq 0.
\label{3.10}
\end{equation}%
\emph{\ Let }%
\begin{equation*}
\beta _{0}\left( x,t\right) =w\mid _{\partial _{1}\Phi },\beta _{1}\left(
x,t\right) =\partial _{x_{1}}w\mid _{\partial _{1}\Phi },g\left( x\right)
=w\left( x,0\right) ,x\in \Omega .
\end{equation*}%
\emph{Denote }%
\begin{equation}
F=\left\Vert \beta _{0}\right\Vert _{H^{1}\left( \partial _{1}\Phi \right)
}+\left\Vert \beta _{1}\right\Vert _{L_{2}\left( \partial _{1}\Phi \right)
}+K.  \label{3.13}
\end{equation}%
\emph{\ Assume that an upper bound }$C_{2}=const.>0$ \emph{\ for the norm }$%
\left\Vert w\right\Vert _{C^{1}\left( \overline{Q}\right) }$ \emph{is known,}%
\begin{equation}
\left\Vert w\right\Vert _{C^{1}\left( \overline{\Phi }\right) }\leq C_{2}.
\label{3.12}
\end{equation}%
\emph{Then there exists a sufficiently small number }$\delta _{0}=\delta
_{0}\left( L,RG_{3/4}\right) \in \left( 0,1\right) $ \emph{dependent only on
listed parameters,\ such that\ if }$F\in \left( 0,\delta _{0}\right) $,
\emph{then} \emph{the following logarithmic stability estimate is valid}%
\begin{equation}
\left\Vert g\right\Vert _{L_{2}\left( \Omega \right) }\leq \frac{M_{1}C_{2}}{%
\sqrt{\ln \left( F^{-1}\right) }},  \label{3.131}
\end{equation}%
\emph{In particular, in the case of IP2, let Assumption holds and (\ref{1.5}%
) be valid. Assume that, in addition to the above, (\ref{1.143}) is valid,
and the number }$\delta $ \emph{in (\ref{1.143}) is so small that }$%
\widetilde{C}\delta \in \left( 0,\delta _{0}\right) $, \emph{where }$%
\widetilde{C}=\widetilde{C}\left( P,L,\Phi ,d\right) >0$\emph{\ is the
number from (\ref{3.94}). Also, assume that for a certain }$\alpha \in
\left( 0,1\right) $\emph{\ the upper bound }$C_{3}$\emph{\ of the norm }$%
\left\Vert f\right\Vert _{C^{2+\alpha }\left( \overline{\Omega }\right) }$%
\emph{\ is given, i.e. }$\left\Vert f\right\Vert _{C^{2+\alpha }\left(
\overline{\Omega }\right) }\leq C_{3}$\emph{. Then}%
\begin{equation}
\left\Vert f\right\Vert _{L_{2}\left( \Omega \right) }\leq \frac{M_{1}C_{3}}{%
\sqrt{\ln \left[ \left( \widetilde{C}\delta \right) ^{-1}\right] }}.\text{ }
\label{3.141}
\end{equation}%
\emph{In addition, there exists a number }$\rho =\rho \left(
L_{0},RG_{3/4}\right) \in \left( 0,1/2\right) $\emph{\ such that if }$F\in
\left( 0,\delta _{0}\right) ,$\emph{\ then the following H\"{o}lder
stability estimate is valid}%
\begin{equation}
\left\Vert w\right\Vert _{H^{1,0}\left( D_{1/2}\right) }\leq
M_{1}C_{2}F^{\rho }.  \label{3.132}
\end{equation}

\textbf{Proof}. In this proof $\varepsilon _{0}=\varepsilon _{0}\left(
L,RG_{3/4}\right) \in \left( 0,1\right) $ denotes different sufficiently
small numbers associated with Lemma 3.1. By (\ref{3.2}), (\ref{3.6}), (\ref%
{3.8}) and (\ref{3.91})%
\begin{equation}
\Omega \subset RG_{1/2}\subset R\Phi .  \label{3.14}
\end{equation}%
Using (\ref{3.3}), (\ref{3.4}) and (\ref{3.10}), we obtain
\begin{equation}
\int\limits_{G_{3/4}}\left( w_{t}-Lw\right) ^{2}\varphi ^{2}dxdt\leq
K^{2}\exp \left( \frac{2\cdot 4^{\nu }}{\varepsilon }\right) ,\forall
\varepsilon \in \left( 0,\varepsilon _{0}\right) .  \label{3.17}
\end{equation}

On the other hand, using Lemma 3.1, we obtain for all $\nu \geq \nu
_{0},\varepsilon \in \left( 0,\varepsilon _{0}\right) $%
\begin{eqnarray}
\int\limits_{G_{3/4}}\left( w_{t}-Lw\right) ^{2}\varphi ^{2}dxdt &\geq
&\int\limits_{G_{3/4}}\left( w_{t}-L_{0}w\right) ^{2}\varphi
^{2}dxdt-M_{1}\int\limits_{G_{3/4}}\left( \left\vert w\right\vert
^{2}+w^{2}\right) \varphi ^{2}dxdt  \notag \\
&\geq &C\int\limits_{G_{3/4}}\left( \frac{\nu }{\varepsilon }\left\vert
\nabla \overline{w}\right\vert ^{2}+\frac{\nu ^{4}}{\varepsilon ^{3}}\psi
^{-2\nu }\overline{w}^{2}\right) \varphi ^{2}\left( x,t\right)
dxdt-M_{1}\int\limits_{G_{3/4}}\left( \left\vert w\right\vert
^{2}+w^{2}\right) \varphi ^{2}dxdt  \notag \\
&&-\frac{C\nu ^{3}}{\varepsilon ^{3}}\exp \left( \frac{2\cdot 4^{\nu }}{%
\varepsilon }\right) \left( \left\Vert \beta _{0}\right\Vert _{H^{1}\left(
\partial _{1}\Phi \right) }^{2}+\left\Vert \beta _{1}\right\Vert
_{L_{2}\left( \partial _{1}\Phi \right) }^{2}\right)  \label{3.18} \\
&&-\frac{C\nu ^{3}}{\varepsilon ^{3}}\left( \frac{4}{3}\right) ^{2\nu }\exp %
\left[ \frac{2}{\varepsilon }\left( \frac{4}{3}\right) ^{\nu }\right]
\int\limits_{\partial _{2}G_{3/4}}\left( w^{2}+\left\vert \nabla
w\right\vert ^{2}+w_{t}^{2}\right) d\sigma .  \notag
\end{eqnarray}%
Fix a number $\nu \geq \nu _{0}\left( L_{0},RG_{3/4}\right) >1$ such that%
\begin{equation}
\left( \frac{5}{6}\right) ^{\nu }<\frac{1}{2}.  \label{3.181}
\end{equation}%
If necessary, decrease $\varepsilon _{0}=\varepsilon _{0}\left(
L,RG_{3/4}\right) \in \left( 0,1\right) $, so that $M_{1}<C\nu /\left(
2\varepsilon \right) ,\forall \varepsilon \in \left( 0,\varepsilon
_{0}\right) .$ Then (\ref{3.18}) leads to the following estimate for all $%
\varepsilon \in \left( 0,\varepsilon _{0}\right) $
\begin{eqnarray*}
\int\limits_{G_{3/4}}\left( w_{t}-Lw\right) ^{2}\varphi ^{2}dxdt &\geq &%
\frac{C}{\varepsilon }\int\limits_{G_{3/4}}\left( \left\vert \nabla
w\right\vert ^{2}+w^{2}\right) \varphi ^{2}\left( x,t\right) dxdt \\
&&-\frac{C}{\varepsilon ^{3}}\exp \left( \frac{2\cdot 4^{\nu }}{\varepsilon }%
\right) \left( \left\Vert \beta _{0}\right\Vert _{H^{1}\left( \partial
_{1}\Phi \right) }^{2}+\left\Vert \beta _{1}\right\Vert _{L_{2}\left(
\partial _{1}\Phi \right) }^{2}\right) \\
&&-\frac{C}{\varepsilon ^{3}}\exp \left[ \frac{2}{\varepsilon }\left( \frac{4%
}{3}\right) ^{\nu }\right] \int\limits_{\partial _{2}G_{3/4}}\left(
w^{2}+\left\vert \nabla w\right\vert ^{2}+w_{t}^{2}\right) d\sigma .
\end{eqnarray*}%
Combining this with (\ref{3.13}), (\ref{3.12}) and (\ref{3.17}) and
decreasing $\varepsilon _{0},$ if necessary, we obtain
\begin{equation}
\int\limits_{G_{3/4}}\left( \left\vert \nabla w\right\vert ^{2}+w^{2}\right)
\varphi ^{2}\left( x,t\right) dxdt\leq C\exp \left( \frac{2\cdot 5^{\nu }}{%
\varepsilon }\right) F^{2}+CC_{2}^{2}\exp \left[ \frac{2}{\varepsilon }%
\left( \frac{5}{3}\right) ^{\nu }\right] ,\forall \varepsilon \in \left(
0,\varepsilon _{0}\right) .  \label{3.19}
\end{equation}%
On the other hand, by (\ref{3.7})%
\begin{equation*}
\int\limits_{G_{3/4}}\left( \left\vert \nabla w\right\vert ^{2}+w^{2}\right)
\varphi ^{2}\left( x,t\right) dxdt\geq \int\limits_{G_{1/2}}\left(
\left\vert \nabla w\right\vert ^{2}+w^{2}\right) \varphi ^{2}\left(
x,t\right) dxdt\geq \exp \left[ \frac{2^{\nu +1}}{\varepsilon }\right]
\int\limits_{G_{1/2}}\left( \left\vert \nabla w\right\vert ^{2}+w^{2}\right)
dxdt,\forall \varepsilon \in \left( 0,\varepsilon _{0}\right) .
\end{equation*}%
Combining this with (\ref{3.19}), we obtain%
\begin{equation*}
\int\limits_{G_{1/2}}\left( \left\vert \nabla w\right\vert ^{2}+w^{2}\right)
dxdt\leq C\exp \left( \frac{2\cdot 5^{\nu }}{\varepsilon }\right)
F^{2}+CC_{2}^{2}\exp \left[ -\frac{2^{\nu +1}}{\varepsilon }\left( 1-\left(
\frac{5}{6}\right) ^{\nu }\right) \right] ,\forall \varepsilon \in \left(
0,\varepsilon _{0}\right) .
\end{equation*}%
Hence, using (\ref{3.181}), we obtain
\begin{equation}
\int\limits_{G_{1/2}}\left( \left\vert \nabla w\right\vert ^{2}+w^{2}\right)
dxdt\leq C\exp \left( \frac{2\cdot 5^{\nu }}{\varepsilon }\right)
F^{2}+CC_{2}^{2}\exp \left( -\frac{2^{\nu }}{\varepsilon }\right) ,\forall
\varepsilon \in \left( 0,\varepsilon _{0}\right) .  \label{3.20}
\end{equation}%
By (\ref{3.3}) and (\ref{3.6}) $G_{1/2}\subset \left\{ t\in \left(
\varepsilon /2,3\varepsilon /2\right) \right\} .$ Hence, the mean value
theorem, (\ref{3.14}) and (\ref{3.20}) imply that there exists a number $%
t^{\ast }\in \left( \varepsilon /2,3\varepsilon /2\right) $ such that for
all $\varepsilon \in \left( 0,\varepsilon _{0}\right) $
\begin{equation}
\left\Vert w\left( x,t^{\ast }\right) \right\Vert _{L_{2}\left( \Omega
\right) }^{2}\leq \frac{1}{\varepsilon }\left\Vert w\left( x,t^{\ast
}\right) \right\Vert _{H^{1}\left( RG_{1/2}\right) }^{2}\leq C\exp \left(
\frac{2\cdot 5^{\nu }}{\varepsilon }\right) F^{2}+CC_{2}^{2}\exp \left( -%
\frac{2^{\nu }}{\varepsilon }\right) .  \label{3.21}
\end{equation}

We have%
\begin{equation*}
w\left( x,t^{\ast }\right) =g\left( x\right) +\int\limits_{0}^{t}w_{t}\left(
x,\tau \right) d\tau .
\end{equation*}%
Hence, using (\ref{3.12}), we obtain
\begin{equation*}
\left\Vert w\left( x,t^{\ast }\right) \right\Vert _{L_{2}\left( \Omega
\right) }^{2}\geq \left\Vert g\right\Vert _{L_{2}\left( \Omega \right)
}^{2}-\varepsilon \left\Vert w_{t}\right\Vert _{L_{2}\left( \Phi \right)
}^{2}\geq \left\Vert g\right\Vert _{L_{2}\left( \Omega \right)
}^{2}-M_{1}C_{2}^{2}\varepsilon .
\end{equation*}%
Combining this with (\ref{3.20}), we obtain%
\begin{equation}
\left\Vert g\right\Vert _{L_{2}\left( \Omega \right) }^{2}\leq
M_{1}C_{2}^{2}\varepsilon +M_{1}\exp \left( \frac{2\cdot 5^{\nu }}{%
\varepsilon }\right) F^{2}+M_{1}C_{2}^{2}\exp \left( -\frac{2^{\nu }}{%
\varepsilon }\right) ,\forall \varepsilon \in \left( 0,\varepsilon
_{0}\right) .  \label{3.22}
\end{equation}%
Choose $\varepsilon =\varepsilon \left( F\right) $ such that%
\begin{equation}
\exp \left( \frac{2\cdot 5^{\nu }}{\varepsilon }\right) F^{2}=\exp \left( -%
\frac{2^{\nu }}{\varepsilon }\right) .  \label{3.23}
\end{equation}%
Hence,
\begin{equation}
\varepsilon =\frac{1}{\ln \left( F^{-2/a}\right) },a=2\cdot 5^{\nu }+2^{\nu
}.  \label{3.24}
\end{equation}%
To ensure that $\varepsilon $ is sufficiently small, i.e. $\varepsilon \in
\left( 0,\varepsilon _{0}\right) ,$ we need to choose $F$ so small that
\begin{equation*}
0<F<\exp \left( -\frac{2}{\varepsilon _{0}}\right) .
\end{equation*}%
Hence, we choose $\delta _{0}=\delta _{0}\left( L,RG_{3/4}\right) =\exp
\left( -2/\varepsilon _{0}\right) .$ Hence, (\ref{3.22}), (\ref{3.23}) and
lead to%
\begin{equation}
\left\Vert g\right\Vert _{L_{2}\left( \Omega \right) }^{2}\leq \frac{%
M_{1}C_{2}^{2}}{\ln \left( F^{-2/a}\right) }+M_{1}\left( 1+C_{2}^{2}\right)
\left( F^{2^{\nu +1}}\right) ^{1/a}=\frac{M_{1}C_{2}^{2}a}{2\ln \left(
F^{-1}\right) }+M_{1}\left( 1+C_{2}^{2}\right) \left( F^{2^{\nu +1}}\right)
^{1/a},  \label{3.25}
\end{equation}%
as long as $F\in \left( 0,\delta _{0}\right) .$ Decreasing, if necessary $%
\delta _{0},$ we obtain (\ref{3.131}) from (\ref{3.25}).

We now prove (\ref{3.141}). Let the function $v\in C^{2+\alpha ,1+\alpha
/2}\left( \mathbb{R}^{n}\times \left[ 0,T\right] \right) ,\forall T>0$ be
the solution of the problem (\ref{1.11}), (\ref{1.12}). Recall that by (\ref%
{1.5}) $f\left( x\right) =0$ in $\mathbb{R}^{n}\diagdown \Omega .$ It
follows from the formula (14.6) of \S 14 of Chapter 4 of the book \cite{LSU}
as well as from (\ref{3.91}) that
\begin{equation}
\left\Vert v\right\Vert _{C^{1}\left( \overline{\Phi }\right) }\leq
\left\Vert v\right\Vert _{C^{2+\alpha ,1+\alpha /2}\left( \overline{\Phi }%
\right) }\leq M_{1}\left\Vert f\right\Vert _{C^{2+\alpha }\left( \overline{%
\Omega }\right) }\leq M_{1}C_{3}.  \label{3.26}
\end{equation}%
By (\ref{1.11}) $K=0$ in (\ref{3.10}). Next, since (\ref{3.94}) follows from
(\ref{1.143}), we obtain for the new number $F$ in (\ref{3.13})
\begin{equation}
F:=\left\Vert \overline{\varphi }_{2}\right\Vert _{H^{1}\left( \partial
_{1}\Phi \right) }+\left\Vert \overline{\psi }_{2}\right\Vert _{L_{2}\left(
\partial _{1}\Phi \right) }\leq \widetilde{C}\delta .  \label{3.27}
\end{equation}%
Thus, (\ref{3.12}) and (\ref{3.131}) imply that (\ref{3.141}) follows from (%
\ref{3.26}) and (\ref{3.27}).

Finally, we prove (\ref{3.132}). Using (\ref{3.100})-(\ref{3.1033}) as well
as Lemma 2.2, we obtain similarly with (\ref{3.20})%
\begin{equation*}
\int\limits_{D_{1/2}}\left( \left\vert \nabla w\right\vert ^{2}+w^{2}\right)
dxdt\leq C\exp \left( 2\lambda \cdot 5^{\nu }\right) F^{2}+CC_{2}^{2}\exp
\left( -2^{\nu +1}\lambda \right) ,\forall \lambda \geq \lambda _{1},
\end{equation*}%
where $\lambda _{1}=\lambda _{1}\left( L,RG_{3/4}\right) >1$ is a
sufficiently large number. Hence, we obtain similarly with (\ref{3.23}) and (%
\ref{3.25})
\begin{equation*}
\int\limits_{D_{1/2}}\left( \left\vert \nabla w\right\vert ^{2}+w^{2}\right)
dxdt\leq M_{1}C_{2}^{2}F^{2\rho }.\text{ }\square
\end{equation*}

\section{The Quasi-Reversibility Method (QRM)}

\label{sec:3}

We construct the QRM only for the more difficult case of IP2.\ The case of
IP1 is similar, and it can be derived from \cite{Kl1}. Also, we work in this
section only in 3-d, keeping the same notations as above. The construction
in the 2-d case is similar. Since it was described in subsection 2.1 how to
stably obtain the Neumann boundary condition in the parabolic case for both
IP1 and IP2, we assume now that we have both Dirichlet and Neumann boundary
conditions at $\partial _{1}\Phi ,$
\begin{equation}
v\mid _{\partial _{1}\Phi }=\overline{\varphi }_{2}\left( x,t\right)
,\partial _{x_{1}}v\mid _{\partial _{1}\Phi }=\overline{\psi }_{2}\left(
x,t\right) .  \label{4.2}
\end{equation}%
The QRM means in our case the minimization of the following Tikhonov
functional%
\begin{equation}
J_{\gamma }\left( v\right) =\left\Vert v_{t}-Lv\right\Vert _{L_{2}\left(
\Phi \right) }^{2}+\gamma \left\Vert v\right\Vert _{H^{4}\left( \Phi \right)
}^{2},  \label{4.3}
\end{equation}%
subject to the boundary conditions (\ref{4.2}). In (\ref{4.3}) $\gamma >0$
is the regularization parameter, which should be chosen in accordance with
the level of the error in the data.

The requirement $v\in H^{4}\left( \Phi \right) $ is an over-smoothness. This
condition is imposed to ensure that $v\in C^{1}\left( \overline{\Phi }%
\right) $: because of (\ref{3.12}) and the embedding theorem.\ However, the
author's numerical experience with the QRM has consistently demonstrated
that one can significantly relax the required smoothness in practical
computation, see \cite{KKKN,KPK,KBK} and chapter 6 of \cite{BKbook}. This is
likely because one is not using an overly small grid step size in finite
differences when computing via the QRM.\ Hence, one effectively works with a
finite dimensional space with not too many dimensions. This means that one
can rely in this case on the equivalence of all norms in finite dimensional
spaces. Thus, most likely one can replace in real computations $\gamma
\left\Vert v\right\Vert _{H^{4}\left( \Phi \right) }^{2}$ with $\gamma
\left\Vert v\right\Vert _{H^{2,1}\left( \Phi \right) }^{2}$.

While (\ref{4.3}) is good for computations, to prove convergence of the QRM,
we need to have zero boundary conditions at $\partial _{1}\Phi .$ Assume
that both functions $\overline{\varphi }_{2},\overline{\psi }_{2}\in
H^{2,1}\left( \partial _{1}\Phi \right) .$ Denote%
\begin{eqnarray*}
r\left( x,t\right) &=&\overline{\varphi }_{2}\left( x,t\right) +x_{1}%
\overline{\psi }_{2}\left( x,t\right) =\overline{\varphi }_{2}\left(
\overline{x},t\right) +x_{1}\overline{\psi }_{2}\left( \overline{x},t\right)
, \\
\widehat{v}\left( x,t\right) &=&v\left( x,t\right) -r\left( x,t\right)
,p\left( x,t\right) =-\left( r_{t}-Lr\right) \left( x,t\right) , \\
\widehat{f}\left( x\right) &=&\widehat{v}\left( x,0\right) =f\left( x\right)
-r\left( x,0\right) .
\end{eqnarray*}%
Using (\ref{1.11}), (\ref{1.12}) and (\ref{4.2}), we obtain
\begin{eqnarray}
\widehat{v}_{t}-L\widehat{v} &=&p\left( x,t\right) ,\left( x,t\right) \in
\Phi ,  \label{4.4} \\
\widehat{v} &\mid &_{\partial _{1}\Phi }=0,\widehat{v}_{x_{1}}\mid
_{\partial _{1}\Phi }=0.  \label{4.20}
\end{eqnarray}%
Thus, we have obtained Inverse Problem 3.

\textbf{Inverse Problem 3 (IP3).} \emph{Find the function }$\widehat{f}%
\left( x\right) $\emph{\ for }$x\in \Omega $ \emph{\ from conditions (\ref%
{4.4}), (\ref{4.20}).}

To solve IP3 via the QRM, we minimize the following analog of the functional
(\ref{4.3})
\begin{eqnarray}
\widehat{J}_{\gamma }\left( \widehat{v}\right) &=&\left\Vert \widehat{v}%
_{t}-L\widehat{v}-p\right\Vert _{L_{2}\left( \Phi \right) }^{2}+\gamma
\left\Vert \widehat{v}\right\Vert _{H^{4}\left( \Phi \right) }^{2},\widehat{v%
}\in H_{0}^{4}\left( \Phi \right) ,  \label{4.6} \\
H_{0}^{4}\left( \Phi \right) &:&=\left\{ u\in H^{4}\left( \Phi \right)
:u\mid _{\partial _{1}\Phi }=u_{x_{1}}\mid _{\partial _{1}\Phi }=0\right\} .
\notag
\end{eqnarray}%
Let $\left( ,\right) $ and $\left[ ,\right] $ be scalar products in $%
L_{2}\left( \Phi \right) $ and $H^{4}\left( \Phi \right) $ respectively.\
Let the function $u_{\gamma }\in H_{0}^{4}\left( \Phi \right) $ be a
minimizer of the functional (\ref{4.6}). Then the variational principle
implies that
\begin{equation}
\left( \partial _{t}u_{\gamma }-Lu,\partial _{t}w-Lw\right) +\gamma \left[
u,w\right] =\left( p,w_{t}-Lw\right) ,\forall w\in H_{0}^{4}\left( \Phi
\right) .  \label{4.7}
\end{equation}%
Lemma 3.1 follows immediately from the Riesz theorem and (\ref{4.7}).

\textbf{Lemma 3.1}. \emph{For every function }$p\in L_{2}\left( \Phi \right)
$\emph{\ and every }$\gamma >0$ \emph{there exists unique minimizer }$%
u_{\gamma }=u_{\gamma }\left( p\right) \in H_{0}^{4}\left( \Phi \right) $%
\emph{\ of the functional (\ref{4.6}). Furthermore the following estimate
holds }%
\begin{equation*}
\left\Vert u_{\gamma }\right\Vert _{H^{4}\left( \Phi \right) }\leq \frac{%
M_{1}}{\sqrt{\gamma }}\left\Vert p\right\Vert _{L_{2}\left( \Phi \right) }.
\end{equation*}

The idea now is that if $u_{\gamma }\left( x,t\right) \in H_{0}^{4}\left(
\Phi \right) $ is the minimizer mentioned in Lemma 3.1, then the approximate
solution of IP3 is
\begin{equation}
\widehat{f}_{\gamma }\left( x\right) =u_{\gamma }\left( x,0\right) .
\label{4.8}
\end{equation}%
The question of convergence of minimizers of $\widehat{J}_{\gamma }$ to the
exact solution is more difficult than the existence question of Lemma 3.1.
To address the question of convergence, we need to introduce the exact
solution as well as the error in the data, just as this is always done in
the regularization theory \cite{Bak,BKbook,T}. We assume that there exists
an \textquotedblleft ideal" noiseless data $p^{\ast }\in L_{2}\left( \Phi
\right) $. We also assume that there exists the ideal noiseless solution $%
\widehat{v}^{\ast }\in H_{0}^{4}\left( \Phi \right) $ of the following
problem
\begin{eqnarray}
\widehat{v}_{t}^{\ast }-L\widehat{v}^{\ast } &=&p^{\ast }\left( x,t\right)
,\left( x,t\right) \in \Phi ,  \label{4.9} \\
\widehat{v}^{\ast } &\mid &_{\partial _{1}\Phi }=0,\widehat{v}_{x_{1}}^{\ast
}\mid _{\partial _{1}\Phi }=0.  \label{4.10}
\end{eqnarray}%
Let $\omega \in \left( 0,1\right) $ be a small number, which we regard as
the level of the error in the data. We assume that
\begin{equation}
\left\Vert p-p^{\ast }\right\Vert _{L_{2}\left( \Phi \right) }\leq \omega .
\label{4.11}
\end{equation}

\textbf{Remark 3.1}. For brevity, we work in this section with the parabolic
IP3. Still, Theorem 3.1 can be easily linked with the original hyperbolic
IP2. Indeed, to ensure that $p\in L_{2}\left( \Phi \right) ,$we need $%
\overline{\varphi }_{2},\overline{\psi }_{2}\in H^{2,1}\left( \partial
_{1}\Phi \right) .$ While Assumption implies (\ref{1.14}), which, in turn
guarantees that $\overline{\varphi }_{2}\in H^{2,1}\left( \partial _{1}\Phi
\right) ,$ there is no guarantee that $\overline{\psi }_{2}\in H^{2,1}\left(
\partial _{1}\Phi \right) $ (see (\ref{1.141})). To ensure the latter, we
should replace in Assumption $C^{3}\left( \mathbb{R}^{n}\times \left[ 0,T%
\right] \right) $ with $C^{5}\left( \mathbb{R}^{n}\times \left[ 0,T\right]
\right) .$ In this case (\ref{1.141}) would be replaced with $\left\Vert
\overline{\psi }_{2}\right\Vert _{C^{2+\alpha ,1+\alpha /2}\left( \overline{P%
}_{1}\right) }\leq C\left( P,L\right) \left\Vert \overline{\varphi }%
_{2}\right\Vert _{C^{4+\alpha ,2+\alpha /2}\left( \overline{P}_{1}\right) }.$
The latter means, in turn that the comparison of the functions $p$ with the
exact function $p^{\ast }$ in (\ref{4.11}) would be replaced with the
comparison of the approximate and exact data $\varphi _{2}$ and $\varphi
_{2}^{\ast }$ of IP2. This can be done via a routine procedure by replacing (%
\ref{1.143}) with $\left\Vert \varphi _{2}-\varphi _{2}^{\ast }\right\Vert
_{C^{5}\left( \overline{P}_{T}\right) }\leq \delta e^{bT},\forall T>0.$ In
this case we would have in (\ref{4.11}) $\omega =\omega \left( \delta
\right) .$

Theorem 3.1 establishes the convergence rate of the QRM. Note that an upper
estimate of the exact solution is often assumed to be known in the
regularization theory, also see Remark 2.2.

\textbf{Theorem 3.1}. \emph{Let conditions (\ref{3.2}), (\ref{4.4}), (\ref%
{4.20}) and (\ref{4.11}) be satisfied and the regularization parameter }$%
\gamma $\emph{\ in (\ref{4.6}) is chosen such that }$\gamma =\gamma \left(
\omega \right) =\omega \in \left( 0,1\right) $\emph{. Let the function }$%
u_{\gamma }\in H_{0}^{4}\left( \Phi \right) $\emph{\ be the unique minimizer
of the functional (\ref{4.6}), which is guaranteed by Lemma 3.1. Let the
upper estimate }$Y=const.>0$\emph{\ for the exact solution }$\widehat{v}%
^{\ast }\in H_{0}^{4}\left( \Phi \right) $\emph{\ be known, }$\left\Vert
\widehat{v}^{\ast }\right\Vert _{H^{4}\left( \Phi \right) }\leq Y.$\emph{\
Then there exists a sufficiently small number }$\omega _{0}=\omega
_{0}\left( L,\Phi \right) \in \left( 0,1\right) $\emph{\ such that if }$%
\omega $\emph{\ is so small that }$\omega \sqrt{\left( Y^{2}+1\right) }\in
\left( 0,\omega _{0}\right) ,$\emph{\ then the following logarithmic
convergence rate takes place}%
\begin{equation}
\left\Vert \widehat{f}^{\ast }-f_{\gamma \left( \omega \right) }\right\Vert
_{L_{2}\left( \Omega \right) }\leq \frac{M_{1}Y}{\sqrt{\ln \left( \omega
^{-1}\right) }},  \label{4.111}
\end{equation}%
\emph{where the function }$f_{\gamma \left( \eta \right) }\left( x\right) $%
\emph{\ is defined in (\ref{4.8}) and }$\widehat{f}^{\ast }\left( x\right) =%
\widehat{v}^{\ast }\left( x,0\right) .$\emph{\ In addition, for every }$%
\omega \in \left( 0,\omega _{0}\right) $\emph{\ there exists a number }$\rho
=\rho \left( L,\Phi \right) \in \left( 0,1/2\right) $\emph{\ such that the
following convergence rate takes place }%
\begin{equation}
\left\Vert \widehat{v}^{\ast }-u_{\gamma \left( \eta \right) }\right\Vert
_{H^{1,0}\left( D_{1/2}\right) }\leq M_{1}Y\omega ^{\rho }.  \label{4.112}
\end{equation}

\textbf{Proof}. It follows from (\ref{4.9}) and (\ref{4.10}) that the
function $\widehat{v}^{\ast }$ satisfies the following analog of (\ref{4.7})%
\begin{equation}
\left( \widehat{v}_{t}^{\ast }-L\widehat{v}^{\ast },w_{t}-Lw\right) +\gamma %
\left[ \widehat{v}^{\ast },w\right] =\left( p,w_{t}-Lw\right) +\gamma \left[
\widehat{v}^{\ast },w\right] ,\forall w\in H_{0}^{4}\left( \Phi \right) .
\label{4.12}
\end{equation}%
Let $\widetilde{v}=u_{\gamma }-\widehat{v}^{\ast }\in H_{0}^{4}\left( \Phi
\right) $ and $\widetilde{p}=p-p^{\ast }\in L_{2}\left( \Phi \right) .$
Subtracting (\ref{4.12}) from (\ref{4.7}), we obtain%
\begin{equation*}
\left( \widetilde{v}_{t}-L\widetilde{v},w_{t}-Lw\right) +\gamma \left[
\widetilde{v},w\right] =\left( \widetilde{p},w_{t}-Lw\right) -\gamma \left[
\widehat{v}^{\ast },w\right] ,\forall w\in H_{0}^{4}\left( \Phi \right) .
\end{equation*}%
Setting here $w:=\widetilde{v}$ and using Cauchy-Schwarz inequality and (\ref%
{4.11}), we obtain%
\begin{equation}
\int\limits_{\Phi _{1}}\left( \widetilde{v}_{t}-L\widetilde{v}\right)
^{2}dxdt+\gamma \left\Vert \widetilde{v}\right\Vert _{H^{4}\left( \Phi
\right) }^{2}\leq \omega ^{2}+\gamma \left\Vert \widehat{v}^{\ast
}\right\Vert _{H^{4}\left( \Phi \right) }^{2}\leq \omega ^{2}+\gamma Y^{2}.
\label{4.13}
\end{equation}%
Since $\gamma \left( \omega \right) =\omega \in \left( 0,1\right) ,$ then (%
\ref{4.13}) implies that $\left\Vert \widetilde{v}\right\Vert _{H^{4}\left(
\Phi \right) }\leq Y+1.$ Hence, using again (\ref{4.13}) as well as
embedding theorem, we obtain with the constant $c=c\left( \Phi \right) >0$
depending only on the domain $\Phi $%
\begin{eqnarray}
\left\Vert \widetilde{v}\right\Vert _{C^{1}\left( \overline{\Phi }\right) }
&\leq &cY,  \label{4.14} \\
\int\limits_{\Phi }\left( \widetilde{v}_{t}-L\widetilde{v}\right) ^{2}dxdt
&\leq &\left( Y^{2}+1\right) \omega ^{2},  \label{4.15}
\end{eqnarray}

We now apply Theorem 2.2. Comparing (\ref{4.15}) and (\ref{4.14}) with (\ref%
{3.10}) and (\ref{3.12}) respectively, we set%
\begin{equation}
K:=F:=\omega \sqrt{\left( Y^{2}+1\right) },C_{2}:=cY.  \label{4.16}
\end{equation}%
Therefore, (\ref{4.111}) and (\ref{4.112}) follow from (\ref{4.16}), (\ref%
{3.131}) and (\ref{3.132}). $\square $

\begin{center}
\textbf{Acknowledgment}
\end{center}

This research was supported by US Army Research Laboratory and US Army
Research Office grant W911NF-11-1-0399.

\end{document}